\documentclass{IEEEcsmag}

\usepackage[colorlinks,urlcolor=blue,linkcolor=blue,citecolor=blue]{hyperref}
\usepackage{xspace}

\usepackage{upmath, color, soul}
\usepackage[dvipsnames]{xcolor}

\usepackage[most]{tcolorbox}
\usepackage{multicol}

\tcbset{
    frame code={},
    center title,
    left=0pt,
    right=0pt,
    top=0pt,
    bottom=0pt,
    colback=Apricot!40,
    colframe=white,
    width=\linewidth,
    enlarge left by=0mm,
    boxsep=5pt,
    arc=0pt,outer arc=0pt,
    }

\newcommand\myparagraph[1]{ \vspace{4pt} \noindent \textbf{#1.}}

\newcommand{\systemname}{\emph{TIM}\xspace}
\newcommand{\revision}[1]{{\color{black}{#1}}}
\newcommand{\revisionTwo}[1]{{\color{black}{#1}}}
\newcommand{\revisionThree}[1]{{\color{black}{#1}}}

\begin{document}

\sptitle{Theme Article: Next-generation Mixed-Reality User Experiences}

\title{Design and Implementation of the Transparent, Interpretable, and Multimodal (TIM) AR Personal Assistant}

\author{Erin McGowan, Joao Rulff, Sonia Castelo, Guande Wu, Shaoyu Chen, Roque Lopez, Bea Steers, Iran R. Roman, Fábio F. Dias, Jing Qian, Parikshit Solunke}
\affil{New York University, New York, (NY), 11201, USA}

\author{Michael Middleton, Ryan McKendrick}
\affil{Northrop Grumman Corp, Falls Church, (VA), 22042, USA}

%\author{Chen Zhao}
%\affil{New York University, Shanghai, (SH), 200124, China}

\author{Cláudio T. Silva}
\affil{New York University, New York, (NY), 11201, USA}

\markboth{Next-generation Mixed-Reality User Experiences}{Next-generation Mixed-Reality User Experiences}

\begin{abstract} %150 WORDS MAX, currently 140
\looseness-1 The concept of an AI assistant for task guidance is rapidly shifting from a science fiction staple to an impending reality. Such a system is inherently complex, requiring models for perceptual grounding, attention, and reasoning, an intuitive interface that adapts to the performer's needs, and the orchestration of data streams from many sensors. Moreover, all data acquired by the system must be readily available for post-hoc analysis to enable developers to understand performer behavior and quickly detect failures. We introduce \systemname, \revision{the first end-to-end AI-enabled task guidance system in augmented reality which is capable of detecting both the user and scene as well as providing adaptable, just-in-time feedback. We discuss the system challenges and propose design solutions. 
We also demonstrate how \systemname adapts to domain applications with varying needs, highlighting how the system components can be customized for each scenario.} 
% We comprehensively describe our implementation of this system, including state-of-the-art machine learning models, insightful analyses aimed at understanding human behavior, and 3D scene tracking. We also discuss associated challenges and propose design solutions. We demonstrate the real-world value of this system using tactical field care and copilot training data.
%in addition to feedback from domain experts.

% The concept of a personal AI assistant for task guidance is rapidly shifting from a staple of science fiction to an impending reality. Such a system is inherently complex, requiring models for perceptual grounding, attention, and reasoning, an intuitive  interface which adapts to the needs of the user, and the collection and analysis of data from many wearable sensors. We introduce \textit{TIM}, an end-to-end system for AI-assisted task guidance in an augmented reality environment. \textit{TIM} leverages state-of-the-art machine learning and visualization techniques for tracking the state of the 3D environment, reasoning about changes in that environment with respect to a given task, interacting with the user as well as modeling their behavior, and data provenance and analytics. We demonstrate the real-world value of this system using copilot and paramedic training data in addition to feedback from domain experts. 

\end{abstract}

\maketitle

%Intro
%Despite the introduction is clearly oriented towards AI-assisted task guidance, there is a lack of how statistical AI (machine and deep learning) are nowadays strictly linked to extended reality paradigms, and in particular AR, that was the degree of reality chosen by the authors to implement the TIM system. To this date, authors should better contextualize why the considered ML models are usually deployed in AR, pointing to proper reference that describe their advantages (e.g., [1, 2, 3, 4]). I use this comment to highlight a general lack of references in the introduction section: several affirmative statements around all this section needs to be supported by references, like “AI-assisted task guidance (AITG) systems, intended to guide a user through the proper and efficient execution of tasks, are finally turning this long-term vision into reality.”}
   
\label{sec:introduction} %ERIN

\begin{figure*}
  \includegraphics[width=\textwidth]{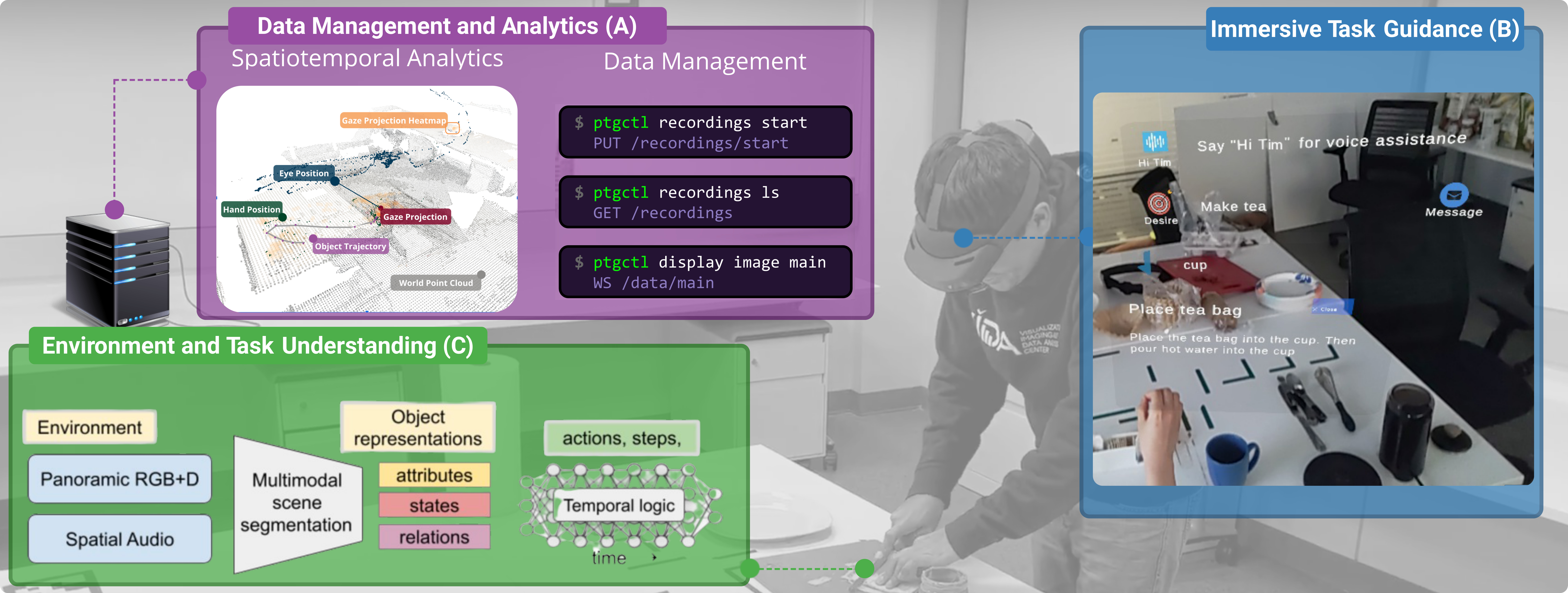}
  \caption{\revision{The main components of the \systemname ecosystem. Tools to facilitate data collection \revisionThree{for} experiment trials and spatiotemporal analysis (A) are crucial to ensure high data quality. Tailored visual widgets to provide feedback to performers and novel interaction mechanisms (B) are needed to ensure smooth guidance throughout different tasks. \revisionThree{S}tate-of-the-art machine learning algorithms \revisionThree{(C)} are needed to perceive the environment and reason about the task's current state.}}
  \label{fig:teaser}
\end{figure*}

\chapteri{I}magine if surgeons had an extra pair of eyes to check their work, or if airplane mechanics could rely on a second brain to guide them through complex repairs on niche vehicles. AI-assisted task guidance (AITG) systems, intended to guide a user through the proper and efficient execution of tasks, are finally turning this long-term vision into reality. %ADD CITATION 
Though the concept of task assistants is not new, the notion that a single system could adapt to guide people with different expertise levels through an infinitude of tasks at different complexity levels has only recently become a fathomable possibility. \revision{These systems are usually mounted in a mixed reality (MR), extended reality (XR), or augmented reality (AR) headset. These have a variety of applications, including ones for facilitating tasks like collaborative manipulation of digital documents \cite{Rev_1_Ref_4}. But} enormous advancements in machine perception and reasoning, along with hardware improvements, have made it possible to begin  \revision{developing a new} generation of \revision{multimodal, AI-enabled task assistants which build intelligent capabilities into the AR system. In 2023, Hirzle et al. \cite{Rev_1_Ref_2} found that (1) Using \revisionThree{artificial intelligence (}AI\revisionThree{)} to understand users (19.3 \%) and (2) Using AI to support interaction (15.4\%) were two of the primary goals of research at the intersection of XR and AI, and that combining XR and AI was beneficial in addressing these questions. Previous works have found AI to be a valuable asset specifically in creating applications in AR as well \cite{Rev_1_Ref_3}.}

\revision{T}his progress is enabling researchers to leverage multimodal data \revision{from} heterogeneous sensors to model tasks, environments, and performers with increasing accuracy. 
\revision{An effective AITG system} must process multimedia streaming data at high rates, employ real-time machine learning (ML) models to understand the physical environment and the \revision{task} performer, and use this information to generate easy-to-understand guidance prompts through different mediums.
These systems can guide the performer using visuals superimposed onto their real-world environment\revisionThree{,} or provide feedback through auditory channels in the form of natural language. \revision{We see this in Stanney et al.'s AI-based XR application for Tactical Combat Casualty Care training \cite{Rev_1_Ref_1}.}  
Furthermore, all data generated during development and testing must be seamlessly persisted in an easy-to-retrieve format\revision{, enabling} retrospective analysis through tailored visualizations that will allow researchers and developers to understand performer behavior and track system failures.

\revision{In this manuscript, we report broadly on \revisionThree{the} multi-year effort of a heterogeneous group of researchers, designers, and developers building \systemname: a Transparent, Interpretable, and Multimodal AR Personal Assistant. Throughout this process, different architectures and modules were tested before defining the one presented in this work. Several specific experiments and case studies are reported in different papers describing various modules of TIM~\cite{wu2024artist, castelo2024hubar, castelo2023argus}. The resulting ecosystem aims to 1) support real-time task guidance and 2) enable the analysis of data collected by the system during task guidance. The next sections depict all pieces of this ecosystem and highlight the motivation for each component, the challenges their implementation imposes, and our proposed design solutions.} 

% we provide a comprehensive description of all the components necessary to build the Transparent, Interpretable, and Multimodal AR Personal Assistant: \systemname. We detail all modules of the \systemname ecosystem, which is comprised of modules serving distinct goals. Each module serves one of two purposes: 1) supporting real-time task guidance or 2) enabling the analysis of data collected by the system during task guidance. Moreover, we highlight the motivation for each component, the challenges their implementation imposes, and our proposed design solutions. 
%
For clarity, we refer to the person guided by \systemname during task execution as the \textbf{performer}, and to the developers and researchers using tools from the \systemname ecosystem to analyze and explore data generated by the system as \textbf{users}.
Our design was inspired by requirements and intermittent feedback from developers of AR systems and domain experts that create and evaluate these systems in the context of the Defense Advanced Research Projects Agency’s (DARPA) Perceptually-enabled Task Guidance (PTG) program \cite{ptg_site}. Our \textit{contributions} are twofold:  \revision{\systemname, the first end-to-end AI-enabled task guidance system in AR which is capable of detecting both the \revisionThree{performer} and scene as well as providing adaptable, just-in-time feedback. \systemname integrates three and a half years of our previous work} in object and action detection, machine reasoning, human-computer interaction \revision{(HCI)}, user modeling, and visual analytics. %citations
We also demonstrate \systemname through two \revision{domain applications which customize components of the system for} real-world data from different domains\revisionThree{:} tactical field care and copilot monitoring.

\section{RELATED WORK}
\label{sec:related-works}

%This section should be similar to the one of the same name in the IEEE VIS 23 ARGUS paper
\subsection{Assistive AR Systems} 
\label{subsec:assistive-AR-systems}

The concept of a personal AI assistant is \revision{familiar to most}; most people carry mobile phones with AI assistant features (e.g. Apple's Siri), or have a personal AI assistant in their homes (e.g. Amazon Alexa, Google Home, etc.). \revision{Such AI assistants are not often integrated into extended reality environments.} While the idea of using AR technologies to build assistive systems that have an internal model of the real world and can augment what a performer sees with virtual content has existed for more than three decades~\cite{caudell_augmented_1992}, \revision{it was not possible to begin effectively implementing such a system until the past decade \cite{Rev_1_Ref_5}.} Advances in AR display technologies and \revisionThree{AI}, in addition to the processing power to run in real time, have enabled this progress. \revision{We see this innovation boom in the development of AR systems for remote collaboration on physical tasks; in a survey of such works published between 2000 and 2018, Wang et al. \cite{Rev_1_Ref_6} found that over 80\% of them were published after 2010. They also found these AR systems were developed for a wide variety of domains, including industry, telemedicine, architecture, \revisionThree{and} teleducation.}

\revision{For task guidance,} previous studies have shown that in-situ instructions provided by assistive AR systems help reduce errors and facilitate procedural tasks~\cite{ockerman2000review, DBLP:journals/cii/FiorentinoUGDM14, uva2018evaluating}. Currently, it is unclear whether assistive AR systems shorten task completion time, as several studies \revision{find} longer times with assistive AR systems~\cite{zheng2015eye} while others find the opposite~\cite{funk2015using}. Nonetheless, most studies agree that AR helps to reduce errors and overall cognitive load \revision{by providing} in-situ instruction and guidance.

AR can be enabled by \revision{various} display technologies, from handheld devices like smartphones and tablets to projector-based solutions and heads-up displays found in airplanes or modern cars.
In this study, we focus on see-through AR head-mounted devices (HMDs), \revisionThree{as} these do not place a significant burden on the performer\revisionThree{,} allow\revisionThree{ing} for free head \revision{and hand} movement. \revision{They} also typically offer a wider range of built-in sensors for modeling the environment and \revisionThree{the} performer such as cameras, microphones, and IMUs.
See-through AR headset displays available today include \revisionThree{the} Microsoft HoloLens~2 (used in our work), Apple Vision Pro, Quest 3, and Magic Leap~2.

%This section should briefly cover the 3D multi-object tracker section from the ISMAR submission “Toward Robust Spatial Perception In-the-Wild:Learning 3D Dynamic Episodic Memory with AR”

\subsection{AI Models to Support Task Guidance} %SHAOYU
\label{subsec:multi-object-tracking}

%\joao{reduce MOT related work and add reasoning. Not clear how MOT fits into the task assistance context}
Several types of AI models must work \revision{together for} effective task guidance. These models typically support an AITG system's ability to perceive the task environment \revision{and performer actions} (perceptual grounding and attention), or \revisionThree{to} reason about how state changes within that environment should influence the guidance conveyed to the performer (reasoning). Multi-object tracking (MOT), which assigns a unique ID to each object of interest, is crucial for maintaining an accurate representation of the state of the task environment. AB3DMOT \cite{Wen:2020:3MT} set a benchmark for 3D MOT, demonstrating that such systems can achieve state-of-the-art performance and operate in real-time. However, most MOT systems encounter limitations in scenarios involving abrupt camera movements, such as in AR task guidance. These systems are also primarily designed and tested using automotive datasets, focusing on tracking a limited set of objects (primarily cars and pedestrians). This may not adequately capture the challenges of AR environments, which usually contain a diverse array of smaller objects.

\revision{Action detection is a crucial complement to object detection when creating a seamless task guidance experience that responds to the performer. Previous works have used deep learning to provide this \textquotedblleft meaningful context-specific feedback" to users performing a task in AR \cite{Rev_1_Ref_7}.} %Nevertheless, prior work in this domain tends to focus on exocentric, or third-person, viewpoints of the performer. This is insufficient for AR task guidance; an egocentric, or first-person, perspective is much more effective as it captures the details of hand-object interactions and performer attention \cite{Rev_1_Ref_8}.}
\revisionTwo{There is much prior work on action recognition using exocentric (third-person) data. However, this approach} is insufficient for AR task guidance. \revisionTwo{We will focus on data captured from an egocentric (first-person) perspective, which} is much more effective as it captures the details of hand-object interactions and performer attention \cite{Rev_1_Ref_8}. \revisionTwo{
The rising popularity of \revisionThree{HMDs} with cameras has led to an increase in egocentric video and, in turn, increased work on the challenge of egocentric action recognition. This challenge is unique from exocentric action recognition in that unpredictable camera movement and lack of context due to a narrow field of view \revisionThree{(FoV)} make recognizing actions more difficult. Previous works have used egocentric action detection to perform tasks that could support AR task guidance. For instance, Lu et al. used egocentric video to automatically break a video into task steps based on hand-object interactions \cite{Lu:2021}. Moreover, Wang et al. were able to enable state-of-the-art action detection using egocentric video alone, without intermediate exocentric transferring \cite{Wang:2023}.} 

%AITG systems also rely on the ability to reason about state changes within this task environment.
\revision{For} machine reasoning, numerous studies \cite{zeng2022socratic} have demonstrated notable performance outcomes across various downstream multimodal applications by integrating parameters from extensive pretrained models and employing multimodal \revisionTwo{end-to-end} joint training. However, most of these systems have not focused on task guidance scenarios from egocentric inputs. \revision{The hand-object interactions and performer attention information conveyed by egocentric inputs allows us to create a system that tailors task guidance to what the performer can see and interact with at \revisionThree{that} moment.} \revisionTwo{Traditional approaches, such as graph-based methods \cite{wang2018active}, \cite{stanescu2023state}, have been successfully applied for task guidance. Unlike our approach, \revisionThree{however,} these methods have primarily been used in scenarios with minimal input and state variation (e.g., in LEGO tasks, where the only objects are the pieces).}

\begin{figure*}
\centering
  \includegraphics[width=0.74\textwidth]{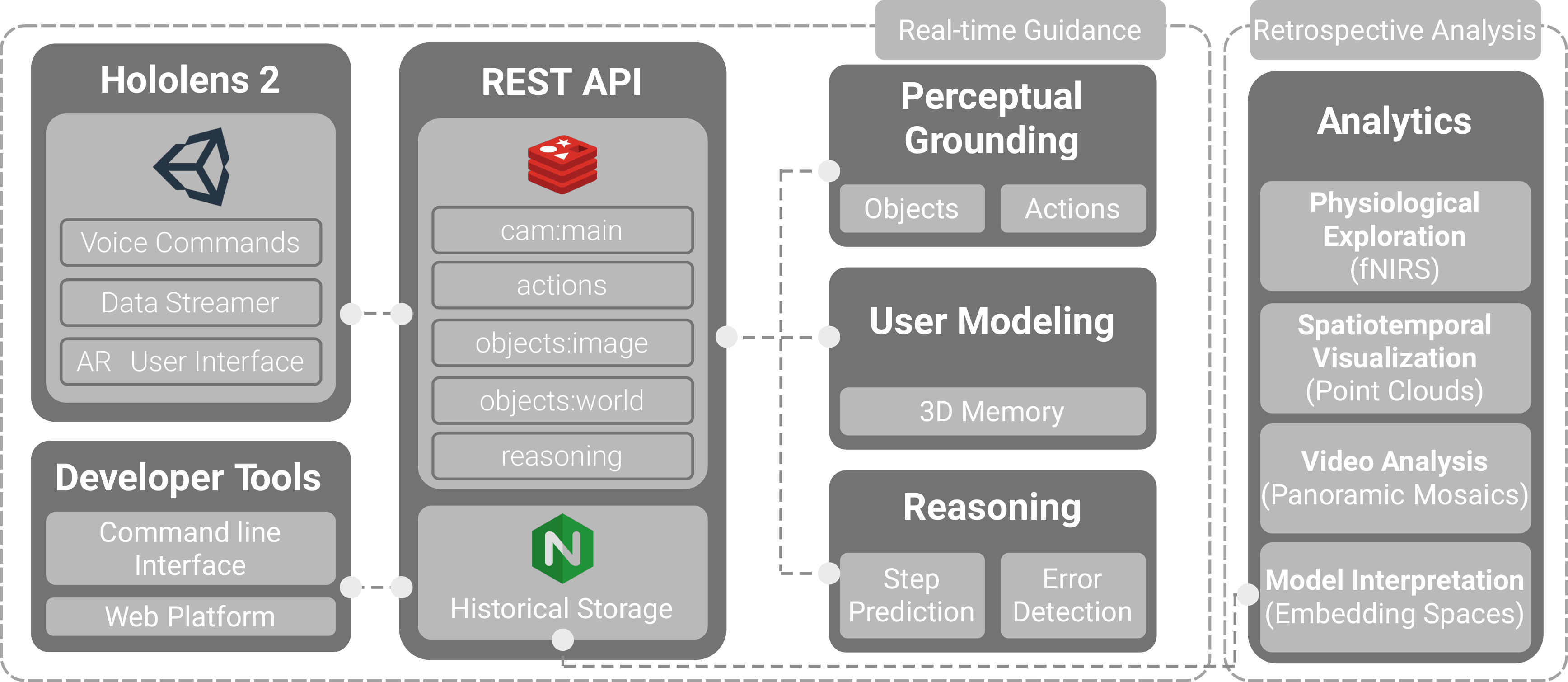}
  \caption{An overview of the \systemname architecture.}
  \label{fig:system_architecture}
  \vspace{-.43cm}
\end{figure*}

%This section should be similar to the section “Visualization of Multivariate Temporal Data” from the IEEE VIS 23 ARGUS paper, but should also (VERY) briefly touch on concepts from the "human behavior analysis based on time series” related works sections from the HuBar paper. 
\subsection{Multimodal Analytics} %MICHAEL
\label{subsec:multimodal-time-series-analysis}

Numerous methods for multimodal temporal visualization tools have been proposed \revisionThree{(e.g.} multiple views, aggregation, level-of-detail\revisionThree{)} \cite{kehrer_visualization_2013,liu_visualizing_2017}. %Aigner_visualization_2011
Recent attempts have \revisionThree{focused on} understanding and debugging  temporal data for multimodal, integrated-AI applications. PSI Studio \cite{bohus_platform_2021} and Foxglove \cite{foxglove} are platforms designed for visualizing multimodal data streams, but \revisionThree{they} require users to format and organize data specifically for their systems. Both primarily facilitate the visualization of data streams rather than summarizing extensive recording periods or debugging ML models. In contrast, Manifold focuses on the interpretation and debugging of general ML models \cite{zhang_manifold_2019}. \revision{We target} a subset of ML models that analyze AI assistant behavior, necessitating different functionalities from existing visualization systems.

One \revisionThree{such} functionalit\revisionThree{y} is human behavioral analysis (HBA), which requires the extraction and correlation of meaningful features from sensor data with human actions. Studies have demonstrated the use of techniques such as time series shapelets to segment behavior activities from sensor data \cite{qin_imaging_2020}. %liu_sensor-based_2015, gong_structured_2014, 
Integrating multiple data streams for a holistic view of behavioral patterns is another important \revisionThree{component} for such analysis \cite{fulcher_feature-based_2017}. One area that is rarely accommodated in HBA visualization is HBA with physiological measures. This is especially sparse in AR tools, where physiological measures are often paired with AR sensor suits to monitor an individual's activity on real world tasks.
 %ERIN, SHAOYU, MICHAEL 
\section{SYSTEM ARCHITECTURE} %JOAO
\label{sec:system-architecture}

One of the biggest challenges \revisionThree{when} developing an AITG system is efficiently coordinating heterogeneous services to process streaming data from \revisionThree{a wide variety of} sensors. 
Each of these services runs complex  algorithms to reason about the environment based on the most recent sensor data. Also, the input for these services is often based on \revisionThree{multiple} data streams that must be synchronized beforehand.
Such a system must follow (among others) three crucial requirements: 1) low latency response times; 2) easy retrieval of multi-source 
streaming data; and 3) seamless persistence of provenance data.

Per these requirements, \systemname's architecture \revision{comprises} four main modules (see Figure \ref{fig:system_architecture}): 1) \revision{the} \textbf{data management module} handles data communication across all modules by utilizing an asynchronous messaging service that any component in \revision{the} system can publish or subscribe to. This is done by Redis Streams, with each stream corresponding to a single camera feed, sensor measurements, or ML model outputs. This module is also responsible for enabling the seamless persistence of data produced in each task session and making them available for download via a static file server. This feature is essential to enable retrospective analytics of system outputs (see \hyperref[sec:data-provenance-and-analysis]{Data Provenance and Analytics}); 2) the \textbf{ML-based models module} groups models responsible for perception and reasoning. All models are deployed in application containers and can retrieve data acquired by the sensors through the data manager; 3) the \textbf{user interface \revision{(UI)} module} sends sensor data and listens for model outputs through the data manager to generate guidance prompts via the AR headset; 4) all data generated during task sessions \revision{is} stored in the \revision{\textbf{data storage module}}, which supports the \systemname analytics tools.

\begin{tcolorbox}[float, floatplacement=b]
\scriptsize
\noindent\rule{\linewidth}{0.4pt}
% SOURCE CODE
The source code for each system component is available on GitHub and linked below.

\noindent\rule{\linewidth}{0.4pt}

% The source code for each component of our system architecture is available on GitHub and linked below.
\begin{itemize}
    \item \href{https://github.com/VIDA-NYU/Perception-training}{Perception}
    \item \href{https://github.com/VIDA-NYU/tim-reasoning}{Reasoning}
    \item \href{ https://github.com/VIDA-NYU/tim-personal-assistant}{AR User Interface}
    \item Data Provenance and Analytics: \href{https://github.com/VIDA-NYU/ARGUS}{ARGUS}, \href{https://github.com/egm68/ARPOV}{ARPOV}, and \href{https://github.com/VIDA-NYU/HuBar}{HuBar}
\end{itemize}
\end{tcolorbox}

% \begin{tcolorbox}
% \scriptsize
% \noindent\rule{\linewidth}{0.4pt}
% % SOURCE CODE
% The source code for each system component is available on GitHub and linked below.

% \noindent\rule{\linewidth}{0.4pt}
% \begin{multicols}{2}
% \begin{itemize}
%     \item \href{https://github.com/VIDA-NYU/Perception-training}{Perception}
%     \item \href{https://github.com/VIDA-NYU/tim-reasoning}{Reasoning}
%     \item \href{https://github.com/VIDA-NYU/tim-personal-assistant}{AR User Interface}
% \end{itemize}
% \columnbreak
% \begin{itemize}
%     \item Data Provenance and Analytics: \href{https://github.com/VIDA-NYU/ARGUS}{ARGUS}, \href{https://github.com/egm68/ARPOV}{ARPOV}, and \href{https://github.com/soniacq/HuBar}{HuBar}
% \end{itemize}
% \end{multicols}
% \end{tcolorbox} %JOAO
%This section should describe both the cooking data we acquired in our own lab environment and the Epic-Kitchens Object States Dataset.

\section{DATA DESIGN AND ENGINEERING} 
\label{sec:domains}

The evolution of the data used to develop \systemname can be described in three phases: initial development, ML model development, and domain adaptation. During initial development, we needed data we could efficiently produce ourselves in a lab environment. This data \revision{also} needed to be collected during a physical task (rather than one performed on a screen) with discrete steps to mimic the \revision{type of} task our industry partners performed in their respective domains. So we chose cooking as our task, performing simple recipes that did not require a stove or oven (e.g. tea, coffee, oatmeal) while wearing a Microsoft HoloLens 2 headset. \revision{The Hololens 2 is \revisionThree{an MR} device \cite{speicher2019mixed} \revisionThree{which includes} sensors to support spatial understanding and sophisticated interactions. Using this headset, we captured egocentric RGB and depth video, audio, hand pose, head position, eye gaze, and IMU data at each time step for each task session. During task execution, all visuals were superimposed onto the real world as \revisionThree{they would be with a} classic AR device.}

When refining our ML models, however, it became clear we needed more data than we could reasonably produce ourselves. We turned to the Epic Kitchens-100 (EK100) dataset \cite{Damen2022RESCALING} due to its size and wealth of annotations. \revisionThree{Yet} EK100 alone was also not sufficient for our task due to its lack of object state information. The ability to infer and understand object states and their transformations through human actions is highly important for the advancement of egocentric perception algorithms, as it allows for the deeper integration of visual data into systems requiring high-level reasoning (such as an AR assistant). To fulfill this need, we augmented EK100 to include object segmentations across all short action clips. These segmentations track the object(s) associated with the action as well as those in the background for the duration of that action. We also defined Planning Domain Definition Language for 57 of 97 verbs in EK100, allowing us to associate objects with their specific states during the periods preceding and following an action. These enhancements enabled us to hone our models for perceptual grounding, attention, and reasoning. 

Finally, we tested our system on real-world data collected by domain experts \revisionThree{(}see \hyperref[sec:domain-applications]{Domain Applications}\revisionThree{)}. 
 %ERIN
    % This section should describe the current perception system, with an emphasis on the ways it improves upon SOTA models for our particular task. It should also include a figure. 

\section{PERCEPTUAL GROUNDING, ATTENTION, AND REASONING} %BEA, IRAN, FABIO, ROQUE
\label{sec:perception}

\subsection{Perception}

\revisionThree{Action recognition} is an important component of egocentric perception systems.
However, systems struggle to deal with different environmental conditions, object occlusion, and overlap of task steps.
To deal with these challenging scenarios, we developed two approaches \revision{that are described and evaluated in more detail in the Supplementary \revisionThree{M}aterial}: 
\emph{(1)} we employed well-suited models to extract action, image (object and scene), and sound features.
These features \revisionThree{were} projected in embedding spaces and combined to feed a recurrent neural network (RNN) that predicts task steps;
\emph{(2)} 
we systematically collected and annotated a series of videos from different object states that are relevant to the proposed case studies and embedded the videos in a feature space.
The states annotated are closely related to each task step.
We also used off-the-shelf models to detect, track, and embed the tracked objects in the space of our collected object state.
Last, we trained a classifier on the object states and used it to predict the states of tracked objects on a video.
With the first approach, we leverage the extractors' capability to represent complex scenes and their components and the RNN\revisionThree{'s} capability to retain past information and use it to predict incoming events.
With the second approach, we focus on the object states and their relations to actions; therefore it is possible to detect many different states in a scene and infer the action associated with each state, enabling the identification of overlapped actions.

%This subsection should provide an overview of the 3D memory system. It should also include a figure. 
\subsection{3D Memory} %SHAOYU
\label{subsec:3D-memory}

3D memory is a critical component that simulates human episodic memory by tracking objects in 3D space. Due to the limited FOV of the HoloLens camera, objects may exit the camera’s view when the \revisionThree{performer} moves their head, rendering them undetectable by the perception module. However, unless physically moved by the \revisionThree{performer}, these objects should remain stationary \revisionThree{in the system's memory}. Thus, although the 2D bounding box positions of objects may shift significantly as the \revisionThree{performer} turns their head, their 3D coordinates should stay constant. By leveraging the capability to memorize and track objects using their 3D world coordinates, AI assistants can offer functionalities unattainable with mere 2D perception. These include guiding \revisionThree{performers} to objects even when they are outside the camera's FOV and utilizing comprehensive object data within the dynamic 3D environment, as opposed to relying solely on objects currently visible. This 3D memory also enables the \revision{UI} to display information and instructions near objects in AR, anchored to their 3D world coordinates. We developed a 3D memory system \cite{10108694} using a hybrid 2D-3D approach that harnesses both the 2D perception in the previous section and the 3D sensing capabilities of the HoloLens. For each object observed, the 3D memory maintains a tracklet that includes data such as the object ID, object class, and 3D positions.
 %FABIO, BEA, IRAN, SHAOYU
% This section should describe the current reasoning system, with an emphasis on how it addresses the unique challenges posed by task guidance (many multimodal data streams to potentially work with, the need to identify not only objects and actions but more ambiguous “steps”). It should also include a figure.

\subsection{Reasoning} %ROQUE
\label{sec:reasoning}
Based on the outputs generated by the percept\revisionThree{ual} grounding and attention modules \revisionThree{(}i.e. object descriptions and states for each frame\revisionThree{)}, the reasoning module implements two approaches to output natural language descriptions of the \revisionThree{current step of the} inferred task and the instructions for the next step. The first method utilizes a dependency graph, while the second employs a random forest model. The outputs are sent to the AR \revision{UI} module for user interactions and the data provenance and analytics module for online and offline analysis. 
%Figure \ref{fig:reasoning} shows an overview of this module.
 
% Source image here: https://docs.google.com/presentation/d/1_Ptax5oRkpH3Vw8Edb62VRuzysqIbNjREHuew3lmIhM/edit#slide=id.g2cf087f234f_0_0

% \begin{figure}
%     \includegraphics[width=\linewidth]{figs/reasoning_workflow.png}
%     \caption{Overview of the reasoning module. At each frame, it outputs the current state with the error status and predicts the next step.}
%     \label{fig:reasoning}
% \end{figure}

%Had to move the UI figure here so it would show up on the right page. Please leave it here. -Erin
% \begin{figure}[!t]
%     \includegraphics[width=\linewidth]{figs/UI.png}
%     \caption{AR GUI. The step number with a progress bar is displayed in the top-left and an image of the step is at the center. The steps cycle appropriately as the user moves through the recipe. To manually move to the previous or next step, buttons are provided below the center image. Objects required for the current step are labeled in blue above the object.
%     }
%     \label{fig:AR_user_interface}
% \end{figure}

In the dependency graph approach,  nodes \revisionThree{represent} task steps, and edges \revisionThree{represent} object states. These object states are `goals’ to be achieved to proceed to the next step. \revision{Each object state is encoded as a vector that includes key attributes of the objects, such as their status and position. The graph is constructed dynamically based on the specific task, ensuring that each step can only proceed when the corresponding object state is satisfied.} For instance, completion of the cooking step `\textit{spread nut butter onto tortilla}' is indicated by achieving the object state `\textit{tortilla-with-nut-butter}.' Additionally, the dependency graph facilitates error monitoring by validating the dependencies of each step, enabling the detection of missing steps or those performed in altered orders.

In the random forest approach, we integrate hand-object interaction \revisionThree{data alongside object states}. Leveraging EgoHOS \cite{zhang2022fine}, we predict the objects the \revisionThree{performer} interacts \revisionThree{with} during the task. The EgoHOS \revisionThree{outputs} serve as feature vectors within the random forests. \revision{These vectors capture whether an object has been manipulated by the right hand, left hand, or both hands, as well as the level of interaction (direct and indirect).} Moreover, this model incorporates object state vectors as additional features. Subsequently, by considering these comprehensive features, the random forest model predicts the ongoing task step.

\revision{Compared to other approaches mentioned in the \hyperref[sec:related-works]{Related Work} section, our module presents several practical advantages. First, the use of graph dependencies provides a clear and interpretable representation of task logic, making the system easy to understand and debug. Additionally,  graph-based methods also ensure predictable and deterministic behavior, which is essential for tasks requiring high reliability. On the machine learning side, the random forest model enhances the module's robustness, as it effectively handles noise and outliers by averaging errors across multiple decision trees. Furthermore, unlike deep learning methods that typically require large datasets for effective training, random forests can achieve strong performance even with a smaller amount of labeled data.}

%\revisionTwo{Details of the preliminary results and the evaluation dataset are provided in the supplementary material.} %ROQUE
% This section should describe the AR user interface and dialogue system, with a focus on any features  that make progress towards the goal of creating a system that adapts to the user’s needs. It should also include a figure. 

\section{AR USER INTERFACE} %GUANDE, JING
\label{sec:AR-UI}
% The AR user interface provides in-situ instructions to support the user's task completion. 

The AR interface provides seamless, responsive, and adaptive task guidance. Our AR interface contains two primary \revisionThree{components}: 1) a stationary, always-on 2D interface for vital information and 2) an adaptive, multimodal interface that uses AI to assist performers in real time. \revision{Inspired by Wu~\cite{wu2024artist},} this interface \revisionThree{analyzes} the performer's current spatial context \revisionThree{and} dynamically simplif\revisionThree{ies} the instructions \revisionThree{where needed}, recognizing what the performer is doing and providing \textbf{relevant guidance} \revision{from the reasoning model described in the previous \revisionThree{section}. This way non-relevant information is filtered \revisionThree{so the} FoV \revisionThree{contains} only important AR instructions, potentially reducing the \revisionThree{performer's} cognitive load~\cite{lindlbauer2019context}. As a result, the AR interface adaptively provides step and guidance} information with the goals of reducing the cognitive burden and assisting with task completion.

%The step number with a progress bar is displayed in the top-left and an image of the step is at the center. The steps cycle appropriately as the user moves through the recipe. To manually move to the previous or next step, buttons are provided below the center image. Objects required for the current step are labeled in blue above the object.
\subsection{HUD Interface}
On the performer's view, our system renders a heads-up display (HUD) showing information required for task completion, such as system commands for controlling the tasks, \revisionThree{the} performer's current step against total steps, task names, and a status bar for the voice assistant (see Figure \ref{fig:teaser}). \revision{The steps cycle appropriately as the \revisionThree{performer} moves through the task. Buttons are provided to move to the previous or next step manually. Objects required for the current step are labeled in blue.} 
\revisionTwo{The objects are detected using the pre-trained zero-shot object detection model Detic~\cite{Zhou:2022}, with a manually crafted prompt that lists potential objects relevant to the task.}
This HUD interface stays with the performer regardless of their location or orientation in 3D\revisionThree{, providing} easy access to vital information. 

Due to limited space, the system collapses task menus during run time, only displaying the active task. Eye tracking enables performers to have hands-free interaction with the task menu (see Figure \ref{fig:teaser}); looking at the task menu expands it, mitigating visual occlusion.
%Should discuss any features that make progress towards the goal of creating a system that adapts to the user’s needs here, including text simplification 
\subsection{Adaptability} 
%While the HUD interface provides basic needs for task guidance, it \revision{also} provides adaptive guidances to performers. These guidances enable the AR system to deal with performer errors and provide a more ``real-person-like'' interactive instruction experience. We provide proactive guidance in the adaptive interface, reducing performers' cognitive load by adapting stationary instructions with the interaction context, making them easier to understand. Overall, the adaptive interface aims to improve \textbf{performance} in addition to providing visual information.
To enable adaptive guidance for context-relevant instructions, we use video streams, \revisionThree{the} performer's AR locations, and physical objects' locations from \revisionThree{the} 3D Memory module (see \hyperref[subsec:3D-memory]{3D Memory}). \revision{For our system, this information helps to create semantically relevant text; for performers, \revision{this information} \revisionThree{can be used to point out (using a floating arrow)} the objects \revisionThree{needed for} the current step.} 

We further developed two different adaptive systems to enrich the AI-supported interaction experience. The first is a text simplification system that reduces the complexity of AR instructions while adding spatial information. This is achieved by using a sequential command to \revisionThree{a large language model (LLM), in our case GPT-3,} to reduce the instructions' complexity, length, and word choice while keeping the meaning intact. Text after simplification will be shorter, but \revisionThree{will} include \revisionThree{information} about physical objects' \revisionThree{locations in relation to the performer}. For example, if the original instruction reads ``place a red cup on top of the machine,'' the simplified instruction \revisionThree{may} read ``place \textbf{the red cup in your left hand} on top of the machine.'' 

The second system is a context-based information guidance system, which uses multimodal LLMs (MLLMs) to analyze the performer's actions, surrounding environment, \revisionThree{and the} tools \revisionThree{they} use, \revisionThree{as well as} surrounding objects' locations, interactivity states, and transformations. For example, \revisionThree{suppose a performer wanted to make a cup of coffee.} The MLLM used in our system is GPT4-oww. The MLLMs \revisionThree{would be} instructed to trace \revisionThree{objects such as the} coffee beans\revisionThree{, and to understand whether the performer is} acting \revisionThree{in accordance with the coffee recipe} instructions. To enable \revisionThree{this} error detection, the task description and common errors are  integrated into the prompt text as the few-shot examples~\cite{castelo2023argus}. The system uses pop-ups, animated tips, and audio to inform adaptive instructions. For animation, we use a set of pre-made, looping animated icons to grab \revisionThree{the} performer's attention. For instance, when the performer encounters hot water \revisionThree{while making the coffee}, a tip with an animated warning icon appears to indicate \revisionThree{they should use caution}; a sound is also played when the performer completes a task. Finally, performers get real-time, multimodal feedback when the system detects deviations from the current task. This multimodal feedback includes a warning message accompanied by voice over asking them to return to their task. 
%\end{itemize}
 %GUANDE, JING
% This section is divided by types of components across all tools (ARGUS, HuBar, ARPOV, VisuARchive) rather than by the tools themselves, since we intend to integrate all of these into ARGUS soon and there’s some overlap in purpose. There is also no specific section for user modeling since it arguably spans all of these, but we should emphasize that we're both analyzing model outputs AND user behavior. 
\section{DATA PROVENANCE AND ANALYTICS}
\label{sec:data-provenance-and-analysis}

%This section should describe the online mode of ARGUS
\subsection{Real-Time Analysis} %SONIA, JOAO
\label{subsec:ARGUS-online-mode}
Real-time debugging is vital for optimizing AR assistan\revisionThree{t} systems, ensuring smooth performance, and user satisfaction \cite{castelo2023argus}. Leveraging our system architecture, which enables seamless streaming data collection and processing, we \revision{have} developed a visual "online mode" \revision{for} instantaneous debugging and validation of AR data. The online mode provides insights into the outputs of reasoning and perception models through tailored visual widgets.

 \begin{figure}
    \includegraphics[width=\linewidth]{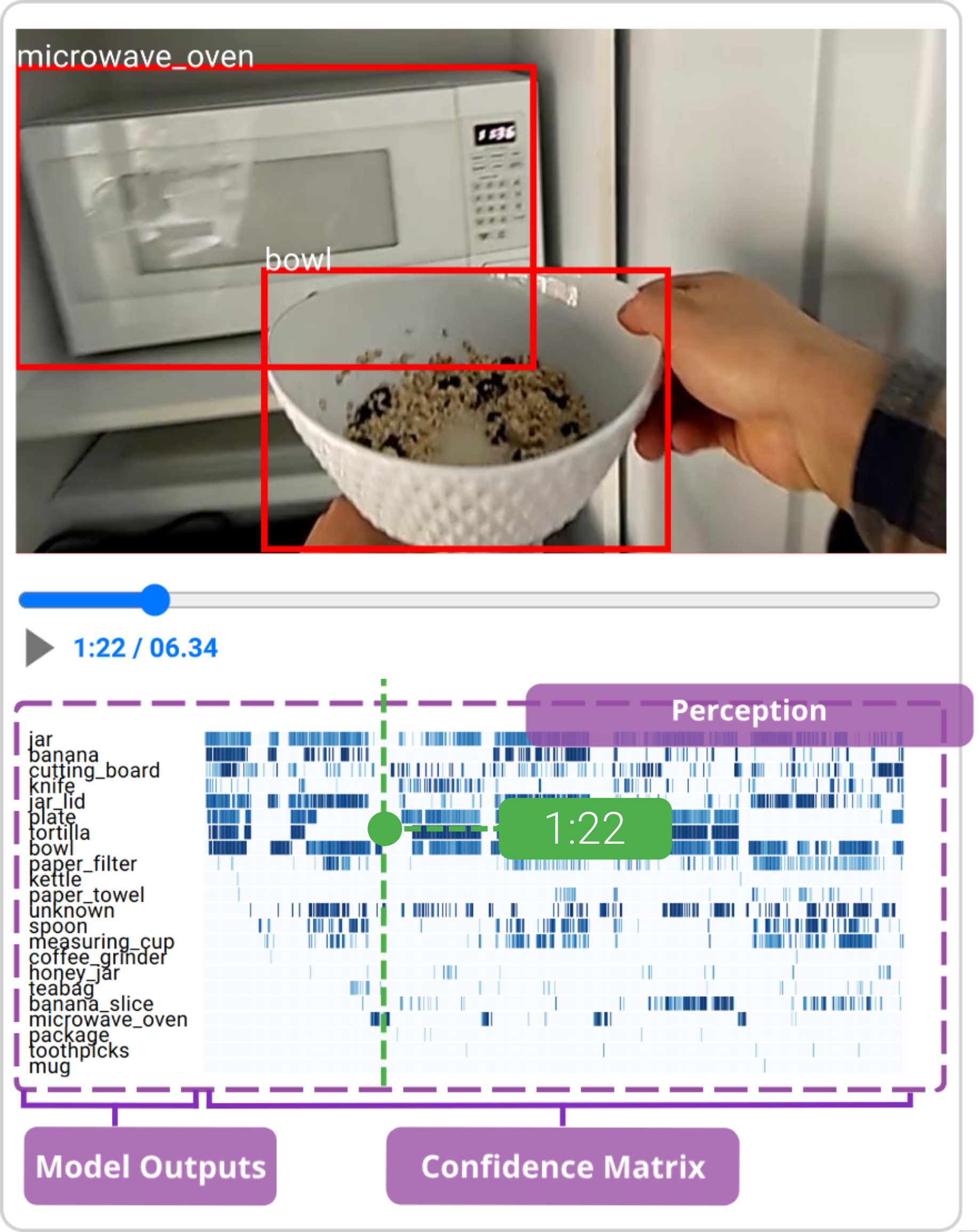}
    \caption{The Model Output View analysis of a cooking session.
    To the left, the model outputs are listed vertically. 
    To the right, the confidence matrix displays the temporal distribution of ML model output confidences across the session. 
    }
    \label{fig:ARGUS_temporal_viewer}
    \vspace{-.2cm}
\end{figure}

\begin{figure*}
\centering
  \includegraphics[width=.95\textwidth]{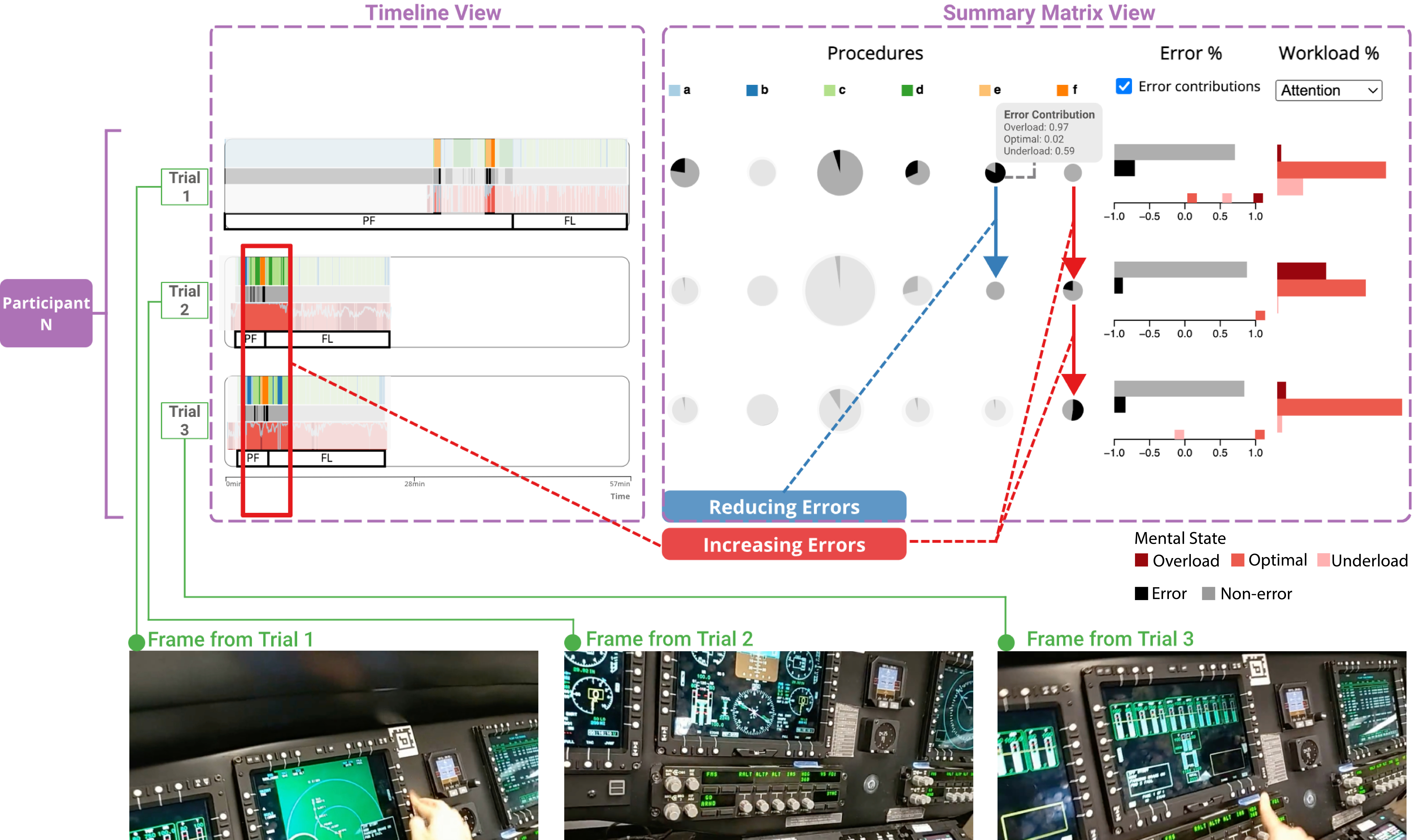}
  \caption{\revision{Timeline View: Performance Overview for Participant N. The Timeline Summary Matrix views depict performance across three consecutive trials under identical task conditions. Key observations include consistent task execution, decreased errors (particularly in Procedure E), increased errors in Procedure F linked to the preflight to flight phase transition, and correlations between errors and mental states. %Notably, during the initial trial, errors in Procedure E correlate with overload mental states, as indicated in the tooltip. 
  Workload summaries demonstrate enhancements in mental states, with the final trial predominantly reflecting optimal states. At the bottom, sample frames from Trials 1, 2, and 3 are displayed.}}
 \label{fig:cs2_enhance_novice_performance}
 \vspace{-.3cm}
\end{figure*}

%This section should describe the Temporal View of ARGUS, the Timeline view of ARPOV (which has overlap with the ARGUS Temporal View), and HuBar. It should also include a figure. 
\subsection{Temporal Analysis} %SONIA, PARI
\label{subsec:temporal-analysis}
ML models are pivotal in AI assistant systems, particularly within the dynamic environment of AR. Despite advancements, tools to enhance their performance are essential~\cite{becher_situated_2022, puladi_augmented_2022}. Model debuggers analyze and refine these models, unveiling insights into their temporal dynamics and decision-making processes. Through temporal analysis, developers can optimize models for accuracy, fairness, and security, enhancing trust in intelligent AR assistants. Our temporal visualizations offer a potent model debugger tailored for AR systems. 
Furthermore, we empower users to explore the temporal distribution of model outputs and \revisionThree{of} data collected from AI-assisted guidance systems\revisionThree{, as well as to} conduct insightful analyses to understand human behavior by leveraging fNIRS data. This expanded capability opens avenues for deeper investigations into the cognitive processes underlying \revisionThree{performer} interactions, ultimately enhancing the design and effectiveness of AR guidance systems for diverse \revisionThree{performer} needs and preferences.

\myparagraph{Model Output View} 
%
% However, the temporal nature of \revision{models for an AR assistant} adds complexity, as they often manage actions or events chronologically. 
%
The Model Output View \revisionThree{facilitates model evlauation} by offering a summary of the temporal distribution of ML model outputs throughout the session (see Figure ~\ref{fig:ARGUS_temporal_viewer}). This is particularly helpful for identifying patterns (e.g. rapid transitions between steps in step detection models) or assessing prediction consistency over time, providing users with a comprehensive overview of model behavior.

% For AR assistant systems, relevant model outputs include objects, actions, and steps.
%
% , which are then used to generate matrix visual representations for temporal model analysis. 
%
The Model Output View contains model outputs, \revisionThree{a} confidence matrix, and global summaries. \revisionThree{\textit{Model outputs} are} grouped by category (e.g. detected objects, \revisionThree{actions, or steps}). The \textit{confidence matrix} displays time on the $x$-axis and confidence scores for detected items in each cell, facilitating detailed analysis. \textit{Global summaries} provide average confidence and detection coverage for each category, enabling quick evaluation. Users \revisionThree{can} explore model outputs at specific times using the temporal controller. 

% Objects, actions, and steps within the confidence threshold are highlighted for further examination, with users able to adjust the threshold using a slider. Object and action labels, along with confidence value bars, are displayed according to guidelines from Felix et al.~\cite{Felix2018TakingWC}.

\myparagraph{Timeline View}
The Timeline View (see Figure \ref{fig:cs2_enhance_novice_performance}, left) facilitates post-hoc analysis of AR task guidance through visualizations highlighting performer behavior, human errors, and cognitive workload responses.

\revision{Four data streams} are visualized \revisionThree{per} selected session, which are organized by trial or subject ID.
Procedures (steps), denoted alphabetically from 'a' to 'f', are shown as horizontal bar graphs colored by procedure and error occurrence. 
%
% To prevent any overlap with the color scale used for the workload variable, we excluded shades of red from the color scale. 
%
Workload status is depicted by segmented bars, with confidence scores shown as a line, illustrating the performer's mental state \revisionThree{with respect to} the chosen workload category over time. Light red segments represent periods of underload, medium red reflects an optimal mental state, and dark red corresponds to \revisionThree{an} overloaded mental state. Finally, sequential data such as flight phase indicators
(see \hyperref[subsec:NGC]{Copilot Monitoring}) are displayed in order.

Temporal alignment of data streams \revision{facilitates} duration evaluation and inter-session comparison of mental states and errors\revision{, as well as} intra-session error identification and correlation analysis.
Users can brush sections to highlight corresponding details in the \textit{Summary Matrix} view (see Figure \ref{fig:cs2_enhance_novice_performance}, right), which complements the timeline by presenting procedure frequencies, error rates, mental state distributions, and correlations between errors and mental states. Pie charts show procedure frequencies and error proportions, with tooltips displaying error-mental state correlations.

%This section should describe the Spatial View of ARGUS and the Panorama View of ARPOV. It should also include a figure. 
\subsection{Spatial Analysis} %JOAO, ERIN
\label{subsec:spatial-analysis}

The temporal analysis of data produced by an intelligent assistant is key to exploring, understanding, and, consequently, improving ML models to support task guidance. 
However, the physical environment where tasks occur often directly influences the output of models supporting guidance. For example, the output of perception models (\hyperref[sec:perception]{Perceptual Grounding, Attention, and Reasoning}) depends \revisionThree{up}on the performer's gaze direction. 
%
% To understand the behavior of all system modules in depth, it is necessary to study system data spatially. Moreover, 
%
Summarizing spatial events can uncover important performer behavior that can guide the development of more adaptive \revision{UI}s based on performer characteristics. 
With this in mind, the \systemname ecosystem includes comprehensive tools to explore the spatial characteristics of data acquired at different scales. 
First, we describe our effort \revisionThree{to} develop intuitive 3D visualizations \revisionThree{which provide an understanding of the performer's} interactions with the physical environment, leverag\revisionThree{ing} depth information acquired by the HoloLens sensors. Second, we present \revisionThree{our} approach to \revisionThree{augmenting the 2D video captured during task performance, expanding the FoV and adding object movement annotations} for a more comprehensive understanding of the scene.

% \begin{figure*}
%     \includegraphics[width=\textwidth]{figs/UI.png}
%     \caption{AR GUI. The step number with a progress bar is displayed in the top-left and an image of the step is at the center. The steps cycle appropriately as the user moves through the recipe. To manually move to the previous or next step, buttons are provided below the center image. Objects required for the current step are labeled in blue above the object.
%     }
%     \label{fig:AR_user_interface}
% \end{figure*}

\myparagraph{3D Visualization} Our approach to visualizing the spatial information captured during task execution aims to facilitate analysis in two ways. First, it \revision{enables} analysis of performer behavior by highlighting their interactions with the physical environment. Second, it enables users to visualize the 3D distribution of model outputs in the physical environment, such as the 3D position\revisionThree{s} of detected objects. 

\begin{figure*}
    \includegraphics[width=\textwidth]{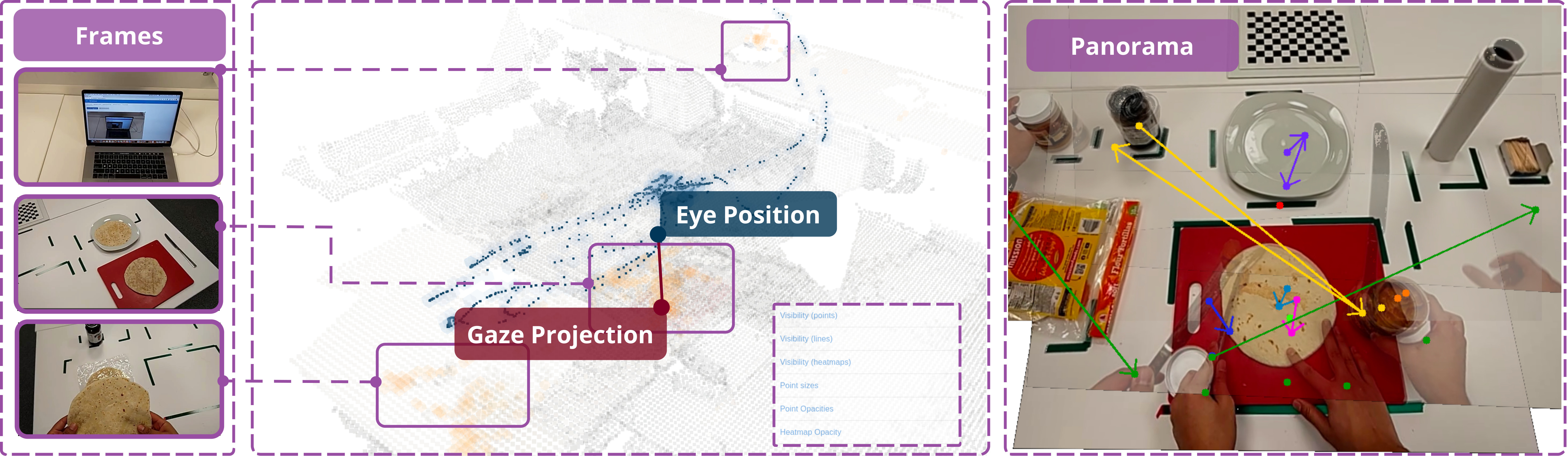}
    \caption{The world point cloud (left) with annotations for spatial data streams and a panorama of selected frames (right) with annotations for object detection model outputs.}
    \label{fig:ARUGS_spatial_viewer}
\end{figure*}

This 3D visualization is \revision{based on} the point cloud representation of the \revision{task} environment\revision{, which} is generated by combining RGB and depth streams captured \revisionThree{by the} HoloLens cameras (see Figure \ref{fig:ARUGS_spatial_viewer}). This representation allows users to understand the physical constraints of the environment and gives context to other data streams, such as performer gaze and position. Users can overlay other data streams onto the scene to gather insights regarding \revisionThree{performer} movement and gaze direction. For example, \revisionThree{we use heatmaps to} denote regions where the performer spent time interacting \revisionThree{with the environment} \revisionThree{(see Figure~}\ref{fig:ARUGS_spatial_viewer}, in orange\revisionThree{)}. \revisionThree{Users can also} hover the mouse over points representing performer position \revisionThree{to} see a ray representing gaze direction.
% The same can be done with 3D model outputs representing the detection of objects in the 3D environment. 
%
This feature \revisionThree{enables} model developers \revision{to} quickly find false positive model outputs by inspecting where specific objects were detected in space.

%This section describes the ARPOV Panorama View
\myparagraph{Augmented Egocentric Video Visualization} Many AR headsets (including the HoloLens 2) contain cameras with a limited FoV that cannot capture everything the performer can see and interact with at a given time. This can hinder analysis of the performance of object detection models using standard methods (i.e. bounding boxes overlaid on a video). We address this issue by allowing the user to select frames from 2D video captured during task performance and generate a panoramic mosaic of those frames. This panoramic mosaic is overlaid with arrows denoting the trajectory of objects detected within the scene, with arrow color corresponding to the detected object label (see Figure \ref{fig:ARUGS_spatial_viewer}). This representation \revision{provides a} more comprehensive understanding of a selected area of the scene, including object positions and perception model failures. For instance, we see in Figure \ref{fig:ARUGS_spatial_viewer} that the perception model correctly detects the performer's hand movement (dark green), but confuses the jar of peanut butter (orange) for the jar of jelly (yellow). 

%Had to move the BBN figure here so it would show up on the right page. Please leave it here. -Erin

%This section should describe the documentation features from VisuARchive.
\subsection{Egocentric Recording Documentation } %GUANDE
\label{subsec:documentation}
Video recordings from the HoloLens 2 cameras offer insights into \revisionThree{task} performance by documenting interactions with objects and actions. These recordings enable professionals, like maintenance workers and surgeons, to optimize and refine procedures by analyzing repetitions and identifying variations or deviations in techniques. This can lead to improved procedural manuals, enhanced safety, and better outcomes.

\myparagraph{Machine Learning Pipeline}
We developed a semi-automatic pipeline integrating vision-language models (VLM) and \revisionThree{LLMs} \revision{for HoloLens recording analysis.} The pipeline processes egocentric video by detecting objects and human-object interactions to identify actions. These findings are then segmented into procedural steps using rule-based algorithms and a VLM, including timing each step. \revisionThree{We use} GPT-4 \revisionThree{to} validate these steps for contextual accuracy, and a GAN-based video summarization model \revisionThree{to} extract key highlights. The final output is an XML-structured document detailing each task step with actions, objects, and a descriptive narrative.
% To document the rich information in the recording, we develop a semi-automatical pipeline based on the vision-language models (VLM) and large language models (LLM). The pipeline consumes the egocentric recording captured from the main camera on HoloLens 2.  The video is first analyzed for object detection and the recognition of human-object interactions, which aids in identifying the actions being performed. These initial results are inputted into the procedural segmentation phase, which utilizes rule-based algorithms along with a vision-language model to systematically organize the video into distinct steps and calculate the time spent on each one. Following segmentation, the GPT-4 model is employed to apply common sense reasoning for output validation, ensuring that the actions and objects recognized are contextually appropriate for the task. After this validation, the information is compiled into an XML-structured document. This document encapsulates the task in a series of steps, each annotated with its associated action, the objects involved, and a descriptive narrative, thus providing a clear and structured representation of the task's execution. We also employ a video summarization model based on the generative adversarial network (GAN)~\cite{DBLP:conf/bmvc/WuLS21} to extract the highlights of each step. These highlights capture the performer's key activities and critical steps essential for the task's success.

\myparagraph{Post-Recording Review} The documented video enables task performers to review their recordings and quickly skim through the highlights. We generate a multimodal document that visualizes each step with its corresponding visual frames and textual descriptions, all based on the XML output from the machine-learning pipeline. This document assists performers in understanding their task process without \revisionTwo{requiring them to manually review the entire task recording.}

\myparagraph{Task Performance Evaluation} 
After reviewing the recordings, task performers may wish to self-evaluate their performance and seek insights for improving their skills. We visualize the time spent on each step, enabling performers to identify bottlenecks where excessive time was spent. Since the different steps may not be balanced and some steps naturally require more time, we support comparisons between various recordings in a summary view. This view visualizes and aggregates the time spent on different steps across recordings. By comparing their \revisionThree{recordings} to others, especially professional recordings, task performers can better understand their performance.

% \myparagraph{Procedural Optimization} Besides the task performers themselves, it is crucial for other professionals, especially trainers, to review the recordings to gain a better understanding of the task procedures and optimize them. An illustrative scenario is where different sequences of steps may be acceptable for task completion but result in varying levels of time efficiency. Therefore, we enable users to query the recording collection based on different step sequences. For example, users can query and compare videos where the coffee filter is prepared before grinding the coffee beans versus the inverse order. By comparing various recordings and delving into their summaries, users can gain insights into optimizing the procedure.

 %SONIA, JOAO, PARI, ERIN, GUANDE

\section{DOMAIN APPLICATIONS}
\label{sec:domain-applications}

%\erin{Needs high level description of each domain, our goals in using this system in these domains, and what the challenges are for each domain}

Once we implemented the \revisionThree{above} perceptual grounding, attention, and reasoning models along with an AR-guided \revision{UI} and tools for data provenance and analytics, we were ready to test \systemname on real-world task performance data. Our chosen tasks are physical, and represent domains (tactical field care [TFC] and copilot monitoring [CM]) that could benefit greatly from an AITG system. However, they also each present unique challenges. \revision{For} TFC, an AITG system could improve  performance and cognitive load of medics in a high-stress battlefield environment, but this chaotic environment introduces motion and noise that may confuse perception and reasoning models. \revision{For CM}, an AITG system could capture performance metrics and provide insights into copilot errors, yet the visualization and analysis of data captured in these circumstances are not trivial. In the following sections, we describe each of these domains, their respective challenges, and how we address those challenges through our design of \systemname
\revision{, noting that our findings are qualitative and reflect the conditions specific to our experimental setting.}

%This section should describe the BBN data and how we're using it in a way that conveys the value of our contributions to this domain. It should also include a figure. 
\subsection{Tatical field care} %BEA, IRAN, FABIO
\label{subsec:BBN}

TFC is provided in a battlefield environment with the appropriate cover. Injuries are treated in order of importance concerning time sensitivity and severity (Massive bleeding, Airway, Respiration/Breathing, Circulation, and Hypothermia/Head injuries)~\cite{ems_tfc}, combining effective tactics and medicine.
These services reduce killed-in-action deaths and can be performed by medical personnel, first responders, or non-medical personnel.
These people have to train in a series of procedures (trauma assessment, applying a tourniquet, etc.) with different step quantities and complexities.

\begin{figure}
  \includegraphics[width=\linewidth]{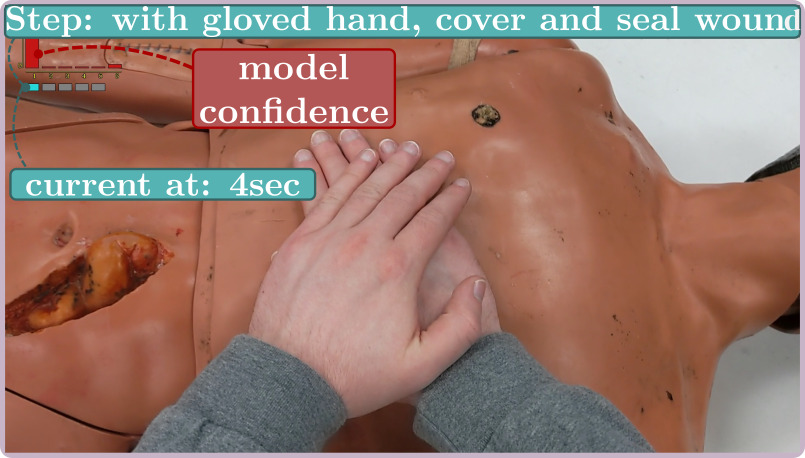}
  \caption{Example of an "apply chest seal" video and \revision{one} its steps identified by our perception approach. }
  \label{fig:bbn_example}
\end{figure}

The training process is performed in different tactical scenarios. Figure~\ref{fig:bbn_example} shows one example of this wherein a trainee is learning to apply a seal to a chest wound while wearing an AR headset capturing egocentric video. 
\revisionThree{The given} frame shows one procedure step and the raw perception outputs \revision{provided by the models described in~\hyperref[sec:perception]{Perceptual Grounding, Attention and Reasoning}.
\revisionThree{See} the Supplementary Material \revisionThree{for} a detailed overview of the dataset used to train and evaluate the models over the TFC tasks}. \revisionThree{Viewing a video with similarly annotated frames,} we see  the model correctly detects each step, though it is most confident in detecting steps \revision{``Cover and seal wound site with hands,'' ``Open vented chest seal package,'' \revisionThree{and} ``Place chest seal with a circle of vents over wound.''} These insights can help ML model developers pinpoint steps, actions, or objects that perception and reasoning models may consistently struggle with within a visually noisy environment.

These insights can be augmented by other \systemname modules; our AR \revision{UI} can guide personnel training, show the procedure progress, and help the \revisionThree{performer} to \revision{both prevent and identify} mistakes (see Figure~\ref{fig:teaser}). \revision{For example, an instruction such as ``Place tourniquet over affected extremity 2-3 inches above wound site'' may encourage the task performer, after completing the step, to read the text carefully and identify potential errors related to the 2-3 inch measurement.}
Furthermore, the \systemname online mode can give insights into the model outputs\revisionThree{,} and the Timeline View tool (see Figure~\ref{fig:ARGUS_temporal_viewer}) can help to assess prediction consistency and give an overview of the models' behavior.

\subsection{Copilot Monitoring} %SONIA, PARI
\label{subsec:NGC}

In the aviation industry, copilots \revisionTwo{possess} varying levels of expertise--from novices to seasoned professionals--\revisionTwo{\revisionThree{and thus require} tailored support to enhance their skills. AR flight guidance systems hold significant potential to address this need by improving performance and overall well-being.

To showcase how the data provenance and analytics modules of our platform can be leveraged to refine AR guidance systems, we present a scenario where a developer evaluates copilots' progress across multiple flight tasks within a mixed-modality AR environment. By identifying performance trends and sources of error, the developer can refine guidance mechanisms to minimize errors and more effectively support pilot training and skill development. Errors were defined as deviations from required actions logged by the mission computer. Details of the data collection protocol are in \cite{castelo2024hubar}.}
%, as documented in the mission computer logs \cite{castelo2024hubar}.}
%
%
% In the aviation industry, copilots have varying levels of expertise, from novices to seasoned professionals. AR flight guidance systems can serve as invaluable tools to assist copilots in improving their performance. This \revision{use case} delves into the iterative process necessary for optimizing AR guidance systems, emphasizing the importance of targeted interventions to enhance both performance and the well-being of copilots.
%
% \revision{We demonstrated how the data provenance and analytics modules of our system can be customized to enhance AR guidance systems for improving copilot performance.} % SONIA
% Consider a scenario where an AR guidance system developer endeavors to evaluate copilots' progress across multiple flight tasks. The objective is to \revision{illustrate how analyzing a mixed-modality AR environment can inform adjustments to guidance systems to better support pilot training over time.}
%
%
%a novice pilot, referred to as "Participant N," 
\revision{``Participant $N$'' \revisionThree{is} a skilled engineer and fast pilot with a background in Blackhawk mission computers. \revisionThree{They c}ompleted the most flights with minimal fatigue\revisionThree{, were} quick in pre-flight procedures\revisionThree{, and were p}aired with Engineer 2, with minimal guidance. This participant}
%who 
undertook the same flight task three times under standard conditions: Trials 1, 2, and 3. %The participant's cognitive workload is measured using fNIRS and classified into three states: optimal (balanced load), overloaded (burden exceeds capacity), and underloaded (insufficient engagement).
\revision{The participant's cognitive workload was measured using Functional Near-Infrared Spectroscopy (fNIRS), a portable and minimally-invasive neuroimaging technique commonly used in pilot studies \cite{ayaz_using_2011, dehais_monitoring_2018}. fNIRS monitors cortical hemodynamics \revisionThree{via} the prefrontal cortex, which is functionally implicated in processing and control of workload facets \cite{fogelson_prefrontal_2009, funahashi_working_1994, petersen_attention_2012}. 
When measuring cognitive workload, it is important that we can determine state changes, which are likely more important than changes in workload levels following minor changes in task demands \cite{mckendrick_deeper_2018}. We \revisionThree{divide} state changes into three categories: optimal (balanced cognitive load), overload (exceeding capacity, hindering new information processing \cite{young_state_2015}), and underload (insufficient engagement, potentially reducing focus \cite{young_state_2015}).
}

\revision{The Timeline View shows} Participant $N$ consistently faced challenges during the preflight phase in all trials (see Figure \ref{fig:cs2_enhance_novice_performance}). However, due to the sporadic occurrence of errors, pinpointing the specific procedures where the copilot struggled most proved difficult. Further examination through the Matrix View unveiled a consistent execution of tasks by Participant $N$ across sessions, with Procedure C emerging as the most prevalent. Notably, substantial errors were observed in Procedures A, D, and E during the initial attempt (Trial 1). %It is worth nothing the the high correlation (0.97) between errors in Procedure E and the "overload" mental state, as indicated in the tooltip visualization. 
Subsequent trials exhibited improvement, notably in Procedures A and E during the second attempt (Trial 2), where errors significantly diminished, especially in Procedure E, dropping from over 70\% to zero. However, errors surfaced in Procedure F during this trial. This trend persisted in the final attempt (Trial 3), with a decline in performance observed in Procedure F but improvements in other procedures.

\revision{T}he Timeline View \revision{shows a} correlation between errors in Procedure F and the transition from the preflight to flight phase, hinting at the necessity for additional guidance during this phase. Furthermore, analyzing mental state through workload summaries revealed a positive impact on the copilot as errors were overcome. \revision{It is important to note that improvements in cognitive workload could be influenced by a learning effect bias due to task repetition, however, repetition is common in pilot training \cite{mumaw_analysis_nodate}.} Despite experiencing high levels of the "underload" mental state in Trial 1, subsequent trials witnessed a decrease in the "underload" mental state, albeit accompanied by an increase in the "overload" mental state in Trial 2. By Trial 3, the copilot achieved an optimal mental state with minimal instances of "underload" and "overload" states.

\revisionThree{This highlights} the connection between overcoming flight errors and \revisionThree{improved copilot} mental state. Recognizing the significance of this, the developer \revisionThree{notes} the need to \revisionThree{enhance} guidance during the transition from preflight to flight, not only to mitigate errors but also to optimize the copilot's mental state.

%  \begin{figure*}
%   \includegraphics[width=\textwidth]{figs/HUBAR_case_study_v4.png}
%   \caption{Timeline View: Performance Overview for Participant N. The Timeline view and Summary Matrix view depict performance across three consecutive trials under identical task conditions. Key observations include consistent task execution, decreased errors (particularly in Procedure E), increased errors in Procedure F linked to the preflight to flight phase transition, and correlations between errors and mental states. %Notably, during the initial trial, errors in Procedure E correlate with overload mental states, as indicated in the tooltip. 
%   Workload summaries demonstrate enhancements in mental states, with the final trial predominantly reflecting optimal states. At the bottom, sample frames from Trials 1, 2, and 3 are displayed.}
%  \label{fig:cs2_enhance_novice_performance}
% \end{figure*} %FABIO, BEA, IRAN, SONIA, PARI
 %Future works will include application to more day-to-day life scenarios and as an assistive device 

\section{CONCLUSION} %ERIN
\label{sec:conclusion}
We presented \systemname, a transparent, interpretable, and multimodal personal assistant for task guidance in AR. We detail the design and end-to-end implementation of \systemname's perceptual grounding, attention, and reasoning models, AR \revision{UI}, and data provenance and analytics capabilities. We also provide two \revision{use cases} \revisionThree{showcasing} how \revision{components of} \systemname have \revisionThree{assisted} domain experts in \revision{TFC} and copilot training applications. 

\revision{One limitation of the \systemname ecosystem is that it can only accommodate physical tasks (rather than ones involving limited movement or performed on a screen). Some components, such as those for data provenance and analysis, require little adaptation between use cases. However, others, particularly the perception and reasoning models, require a more involved customization process. These models also may not perform as expected in environments with different lighting conditions. Moreover, additional efforts are required to support collaboration among multiple \revisionThree{performers} on a task.}

%\erin{We should add a couple of specific future goals here.}
We envision \systemname to unlock several avenues for future research connecting \revision{HCI}, visualization, and ML communities \revision{through} the goal of developing more reliable intelligent AR systems. In the future, we intend to improve the general accuracy of our perception and reasoning models as well as their ability to generalize to other tasks and domains. We also plan to delve deeper into modeling the cognitive workload of the performer, allowing us to further adapt task guidance to the performer's needs. Moreover, while \systemname was designed and tested with complex, physical tasks performed by domain experts, we believe personal AI assistant systems in AR will also be used by the average layperson in day-to-day tasks.
We hope to explore the possibility of adapting \systemname for this purpose.

%The manuscript should include a conclusion. In this section, summarize what was described in your paper. Future directions may also be included in this section. Authors are strongly encouraged not to reference multiple figures or tables in the conclusion; these should be referenced in the body of the paper. %ERIN

\section{ACKNOWLEDGMENTS} 
\revisionThree{We thank Saksham Bassi (Amazon) and Chen Zhao (NYU Shanghai) for their contributions to the development of the components described in \hyperref[sec:reasoning]{Reasoning}. Moreover, we thank Ethan Brewer (Spectral Sciences, Inc.) and Michael Krone (University of Tübingen) for their contributions to the development of the components described in \hyperref[sec:data-provenance-and-analysis]{Data Provenance and Analytics}. We thank Brian VanVoorst (BBN Technologies) for providing the TFC dataset. Finally, we thank Juan Bello, Kyunghyun Cho, He He, Qi Sun, and Huy Vo (all New York University) for their contributions to the overall vision and design of this system. 

This work was supported by the DARPA PTG program. Any opinions, findings, conclusions, or recommendations expressed in this material are those of the authors and do not necessarily reflect the views of DARPA.}  

%\begin{thebibliography}{1}

%\bibitem{AA1}
%G. M. Amdahl, G. A. Blaauw, and F. P. Brooks, ``Architecture of the IBM System/360,'' {\it IBM J. Res. \& Dev}., vol. 8, no. 2, pp. 87--101, 1964. (journal)

%\end{thebibliography}

\bibliographystyle{abbrv-doi} 
\bibliography{references}
%\printbibliography[title={First Bibliography}, keyword=one]

\begin{IEEEbiography}{Erin McGowan} is a computer science (CS) Ph.D. candidate at New York University (NYU), with research interests including data visualization and analytics, machine learning, and human-computer interaction. Contact at erin.mcgowan@nyu.edu. 
\end{IEEEbiography}

\begin{IEEEbiography}{Joao Rulff} is a CS Ph.D. candidate at NYU. His research includes visualization, visual analytics, human-computer interaction, and urban computing. Contact him at jr4964@nyu.edu.
\end{IEEEbiography}

\begin{IEEEbiography}{Sonia Castelo} is a research engineer and CS Ph.D. candidate at NYU. Her research interests include data visualization and analytics, machine learning, and augmented reality. Contact her at s.castelo@nyu.edu.
\end{IEEEbiography}

\begin{IEEEbiography}{Guande Wu} is a CS Ph.D. candidate at NYU. His research interests include human-AI collaboration and visual analytics. Contact him at guandewu@nyu.edu.
\end{IEEEbiography}

\begin{IEEEbiography}{Shaoyu Chen} is a CS Ph.D. candidate at NYU. His research interests include virtual reality and augmented reality. Contact him at sc6439@nyu.edu.
\end{IEEEbiography}

\begin{IEEEbiography}{Roque Lopez} is a research engineer at NYU. His research interests include applied machine learning, natural language processing and reinforcement learning. Contact him at rlopez@nyu.edu.
\end{IEEEbiography}

\begin{IEEEbiography}{Bea Steers} is a research engineer at NYU. Her research interests include acoustic monitoring networks and urban science. Contact her at bsteers@nyu.edu. 
\end{IEEEbiography}

\begin{IEEEbiography}{Iran Roman} is a postdoctoral researcher at NYU's Music and Audio Research Laboratory. His research interests include theoretical neuroscience and machine perception. Contact him at roman@nyu.edu. 
\end{IEEEbiography}

\begin{IEEEbiography}{Fábio F. Dias} is a post-doctoral associate at NYU. His research interests include data visualization and analytics, machine learning, and signal and image processing. Contact him at ffd2011@nyu.edu. 
\end{IEEEbiography}

\begin{IEEEbiography}{Jing Qian} is a postdoctoral researcher at NYU. His research interests include human-computer interaction and human-AI collaboration. Contact him at jq2267@nyu.edu. 
\end{IEEEbiography}

\begin{IEEEbiography}{Parikshit Solunke} is a CS Ph.D. student at NYU. His research interests include Explainable AI (XAI) and Urban Analytics. Contact him at parikshit.s@nyu.edu. 
\end{IEEEbiography}

\begin{IEEEbiography}{Michael Middleton} is an applied AI engineer at Northrop Grumman. His research interests include applied brain-computer interfaces, augmented reality guidance systems, procedural content generation. Contact him at Michael.Middleton@ngc.com.
\end{IEEEbiography}

\begin{IEEEbiography}{Ryan Mckendrick} is an applied cognitive scientist at Northrop Grumman. His research interests include human-machine augmentation, Neuro cognitive state prediction, and surrogate modeling and optimization. Contact him at Ryan.McKendrick@ngc.com.
\end{IEEEbiography}

%\begin{IEEEbiography}{Chen Zhao} is an Assistant Professor of CS at NYU Shanghai. His main research interest is in natural language processing. Contact him at cz1285@nyu.edu.
%\end{IEEEbiography}

\begin{IEEEbiography}{Cláudio T. Silva} is an Institute Professor of Computer Science and Engineering and Data Science at NYU. He is Co-Director of the Visualization, Imaging and Data Analysis Center and an IEEE Fellow. Contact him at csilva@nyu.edu. 
\end{IEEEbiography}

\end{document}

\section{PERCEPTUAL GROUNDING, ATTENTION, AND REASONING} 

\subsection{Perception}

%%REFERENCES ARE IN THE SAME references.bib

The capacity to recognize and identify human actions is an important component of egocentric perception systems.
However, systems struggle to deal with different environmental conditions, object occlusion, and overlap of task steps.
To deal with these challenging scenarios, we developed two approaches: 

\noindent\textbf{in the first approach}, we slide a $k$-second window over each video and extract different feature modalities from them.
As the task is based on action recognition, we employed Omnivore~\cite{Girdhar:2022} as an action-feature extractor because it demonstrates strong generalization across different datasets, achieving high accuracy levels.
Following this, in each window clip we identify a representative frame (the last window frame in our tests) and extract two sets of image-related features: in the first set, called global features, we use 
CLIP~\cite{Radford:2021} image encoder to embed the frame into a feature space; and in the second set we use an Yolo (v8)~\cite{Redmon:2016}, fine-tuned by domain specialists (cf.~\hyperref[subsec:BBN]{Tatical field care}),
to detect important objects for each task.
The objects are also embedded into a CLIP (region features) feature space.
We project global and regional features with linear layers and combine them using a weighted summation. The weights are defined following an idea inspired by the ConceptFusion~\cite{Jatavallabhula:2023} process of computing pixel-aligned features.
In addition to action and image features, we can also extract sound features with models such as SlowFast Auditory~\cite{Kazakos:2021}.
Finally, action, image, and sound features are projected with linear layers and concatenated to feed a Gated recurrent unit (GRU) followed by a linear layer that predicts task steps. 
In our tests, sound features did not help the results because some tasks have no representative sounds or are performed in noisy environments.
Therefore, all reported tests with GRU use only action and image features. 

\noindent\textbf{in the second approach}, we focused on object states instead of task steps.
Initially, we manually identified the main task objects and the object states that change over tasks and characterize different steps.
With this information, we recorded videos in different circumstances (background, lighting, object types, etc.), labeled the videos with object states, cropped the objects from the frames, embedded them with CLIP, 
and trained a kNN with those features to predict the object states.
Following this, we considered a pipeline that could extend the process for streaming videos. 
The pipeline consists of the Detector with image classes (Detic) model~\cite{Zhou:2022} to identify and segment objects and EgoHOS~\cite{Zhang:2022} to segment hands.
The object and hand masks are used as inputs for the XMem model~\cite{Cheng:2022} which tracks the object in a video, ignoring hands that deal with the objects.
With the masks provided by XMem for the whole video, our pipeline extracts the related objects from the video frames and also embeds them with CLIP image encoder.
Finally, our kNN predicts the object states.

With the first approach, we leverage the capability of the extractors to represent complex scenes and their components and the RNN capability to retain past information and use it to predict incoming events.
With the second approach, we focus on the object states and their relations to actions, therefore it is possible to detect many different states in a scene and infer the action associated with each state, enabling the identification of overlapped actions.

During initial development, we needed data that could be recorded in the laboratory environment and contain simple and well-defined tasks, so we chose the cooking domain. \autoref{tab:gru_cooking} presents the accuracy results of GRU (first approach) to identify the steps of three simple cooking recipes.
Our second approach detects object states that work as input to the reasoning process evaluated in~\hyperref[subsec:reason]{Reasoning} section.
%% In the sequel,~\autoref{tab:object_states} reports the accuracy results of object states identification (second approach). As described in the main text, the outputs of this approach feed the reasoning process evaluated in~\hyperref[subsec:reason]{Reasoning} section. 

\begin{table*}[!h]
    \centering  
    \caption{GRU evaluation results of cooking domain}
    \vspace{5pt} % Adds 5pt of space between the caption and the table
    \label{tab:gru_cooking}
    \begin{tabular}{ll}
    \toprule    
    recipe         & accuracy \\
    \midrule              
    Pinwheels & 0.91     \\
    Coffee    & 0.49     \\
    Mugcake   & 0.44    \\
    \bottomrule    
    \end{tabular}
\end{table*}

Once we implemented and evaluated the approaches, we were ready to test on real-world task performance data.
As we considered domains such as those described in~\hyperref[subsec:BBN]{Tatical field care}, it was possible to replicate only the GRU approach because the dataset does not provide object state annotations.
\autoref{tab:gru_bbn} presents a detailed evaluation summary over a test subset extracted from each skill (task) on the dataset.

\begin{table*}[!h]
  \centering  
  \caption{GRU evaluation results of Tactical Field Care domain}
  \vspace{5pt} % Adds 5pt of space between the caption and the table
  \label{tab:gru_bbn}
  \begin{tabular}{lllll}
  \toprule    
   skill  & \multicolumn{1}{c}{precision} & \multicolumn{1}{c}{recall} & \multicolumn{1}{c}{F1-score} & \multicolumn{1}{c}{accuracy} \\
  \cmidrule(lr){1-4}     
  \cmidrule(lr){5-5}  
  M2   & 0.47$\pm$0.23    & 0.62$\pm$0.14 & 0.50$\pm$0.20 & 0.62 \\    
  M3   & 0.72$\pm$0.16    & 0.74$\pm$0.16 & 0.73$\pm$0.15 & 0.74 \\    
  M5   & 0.64$\pm$0.24    & 0.74$\pm$0.23 & 0.63$\pm$0.17 & 0.74 \\    
  R18  & 0.71$\pm$0.15    & 0.77$\pm$0.09 & 0.73$\pm$0.09 & 0.77 \\    
  \midrule
  A8   & 0.64$\pm$0.19    & 0.83$\pm$0.13 & 0.70$\pm$0.10 & 0.83 \\
  M4   & 0.84$\pm$0.14    & 0.86$\pm$0.06 & 0.85$\pm$0.09 & 0.86 \\
  R16  & 0.70$\pm$0.19    & 0.79$\pm$0.15 & 0.74$\pm$0.14 & 0.79 \\
  R19  & 0.60$\pm$0.22    & 0.67$\pm$0.14 & 0.62$\pm$0.16 & 0.67 \\
  \bottomrule  
  \end{tabular}
\end{table*}

\subsection{Reasoning}
\label{subsec:reason}

We conducted preliminary experiments to evaluate the performance of the Random Forest approach on five cooking recipes: Quesadilla, Coffee, Oatmeal, Tea, and Pinwheels. The annotated dataset was relatively small, comprising 34 videos in total (7 Quesadilla videos, 6 Coffee videos, 7 Oatmeal videos, 7 Tea videos, and 7 Pinwheels videos). Due to the limited dataset size, we used leave-one-out cross-validation for evaluation.

Table \ref{tab:rf_cooking} presents the average accuracy values for each cooking recipe in identifying recipe steps. The highest accuracy was achieved for the Quesadilla recipe, while the Pinwheels recipe showed the lowest performance. A possible explanation for this behavior is the difference in the number of steps: the Quesadilla recipe contains 7 steps, whereas the Pinwheels recipe involves 12 steps, which may increase the complexity and difficulty of the classification process.

\begin{table*}[!h]
    \centering  
    \caption{Results for the Random Forest method of cooking domain}
    \vspace{5pt} % Adds 5pt of space between the caption and the table
    \label{tab:rf_cooking}
    \begin{tabular}{ll}
    \toprule    
     recipe         & accuracy \\
    \midrule    
    Quesadilla & 0.78 \\
    Coffee & 0.63 \\
    Oatmeal & 0.58 \\
    Tea    & 0.46   \\
    Pinwheels & 0.42    \\
    \bottomrule    
    \end{tabular}
\end{table*}

\section{DOMAIN APPLICATIONS}

\subsection{Tactical field care}
\label{subsec:BBN}

The Tactical Field Care Dataset was collected by RTX BBN Technologies as part of the PTG program \cite{ptg_site}.
It includes 830 egocentric videos with durations ranging from approximately 16 seconds to 4 minutes.
These videos cover nine different medical interventions, termed ``skills,'' each annotated by specialists with procedural steps; each intervention contains 3 to 13 steps of varying lengths.
There are approximately 53 unique steps across all skills.
The dataset provides a comprehensive overview of these critical medical interventions, detailing each skill through its sequence of steps aimed at achieving specific objectives. 
The dataset includes the following Airway (A), Massive bleeding (M), and Respiration/Breathing (R) skills:

\begin{enumerate}

\item[\textbf{A8}]\textbf{- Nasopharyngeal Airway (NPA):} The objective of this skill is to secure the airway of a casualty using an NPA tube. This skill involves 5 steps: selecting the appropriate tube, positioning the casualty with the head tilted back, lubricating the NPA, inserting it perpendicularly into the nostril until the flange meets the tip of the nose, and assessing airway compliance by looking, listening, and feeling.

% \item[\textbf{M1}]\textbf{- Trauma Assessment:} The objective of this skill is to perform a comprehensive assessment of a trauma casualty. This procedure consists of 13 steps, summarized as: removing clothing (if necessary), sweeping and raking the limbs and torso, rolling over and checking the back, checking for breathing and pulse, assessing skin temperature and quality, exposing and examining the chest, and potentially re-clothing the casualty.

\item[\textbf{M2}]\textbf{- Apply Tourniquet:} The goal of this skill is to control hemorrhage by applying a tourniquet. This procedure involves 8 steps, summarized as: placing the tourniquet 2-3 inches above the wound, tightening it, securing the strap to the tourniquet body, rotating the windlass to control hemorrhage, locking the windlass in place, securing the remaining strap and windlass keeper, and marking the time of application on the strap.

\item[\textbf{M3}]\textbf{- Apply Pressure Dressing:} This skill aims to control bleeding by applying a pressure dressing to the wound. It includes 5 steps: applying direct hand pressure to the wound, opening the dressing packaging, placing the dressing with direct pressure over the wound site, wrapping the dressing with 80\% elastic stretch, and securing it with the provided device.

\item[\textbf{M4}]\textbf{- Wound Packing:} The objective of this procedure is to control bleeding by packing the wound with appropriate material. It consists of 3 steps: applying direct hand pressure to the wound, opening the dressing packaging, and packing the wound with the provided material.

\item[\textbf{M5}]\textbf{- X-Stat:} The goal of this skill is to control severe hemorrhage using an X-Stat device. This procedure involves 5 steps, summarized as: removing the applicator from its packaging, inserting it into the wound track as close to the bleeding source as possible, deploying sponges by pressing the plunger, and applying manual pressure if necessary.

\item[\textbf{R16}]\textbf{- Ventilate with Bag-Valve-Mask (BVM):} The objective of this skill is to provide ventilation to a casualty using a BVM. It includes 5 steps: positioning the casualty with the head tilted back and nostrils exposed, opening the BVM package, attaching and expanding the mask, placing the mask over the casualty’s mouth, and squeezing the BVM while holding the mask in place.

\item[\textbf{R18}]\textbf{- Apply Chest Seal:} The goal of this skill is to seal a chest wound to prevent pneumothorax. This procedure involves 5 steps: applying initial pressure to the wound, opening the chest seal package, cleaning the wound site, and applying the chest seal over the wound, ensuring proper placement of vents.

\item[\textbf{R19}]\textbf{- Needle Chest Decompression:} The objective of this skill is to relieve pneumothorax through needle decompression. This procedure comprises 6 steps: locating the insertion site at the second intercostal space at the midclavicular line, cleaning the insertion site with alcohol, preparing and inserting the needle, leaving it in place for 5-10 seconds before removing it, removing the needle, and securing the catheter with tape.
\end{enumerate}

\printbibliography